\documentclass[pre,aps,twocolumn,showpacs,superscriptaddress]{revtex4}
\usepackage{graphicx} 
\usepackage{amsthm} 
\usepackage{amsmath}
\begin{document}
\newtheorem{theorem}{Theorem}
\newtheorem{lemma}{Lemma}

\title{Spectral Properties and Synchronization in Coupled Map Lattices}
\author{J. Jost}
\email{jjost@mis.mpg.de}
\affiliation{Max Planck Institute for Mathematics in the Sciences\\
        Inselstrasse 22-26, D-04103 Leipzig, Germany}
\affiliation{Santa Fe Institute, 1399 Hyde Park Road, Santa Fe, NM 87501, USA}
\author{M. P. Joy}
\email{mjoy@mis.mpg.de}
\affiliation{Max Planck Institute for Mathematics in the Sciences\\
        Inselstrasse 22-26, D-04103 Leipzig, Germany}
\date{September 1, 2001}
\begin{abstract} Spectral properties of Coupled Map Lattices are described.
Conditions for the stability of spatially homogeneous chaotic solutions are derived
using linear stability analysis. Global stability analysis results are also
presented. The analytical results are supplemented with numerical examples. 
The quadratic  map is used for the site dynamics with different coupling schemes
such as global coupling, nearest neighbor coupling,
 intermediate range coupling,
random coupling, small world coupling and scale free coupling.
\end{abstract}


\maketitle

\section{Introduction}

Synchronization of large interacting systems has been observed in
several natural situations such as synchronized flashing of the fire
flies, pace maker cells of the heart, neurons, etc \cite{sync,gray,gla01}.
Synchronization of chaos in low dimensional systems was studied
by Pecora and Carroll \cite{pecocaro90}.  It has also been studied in
coupled oscillator systems and other spatially extended systems
\cite{mirostrog,heagy94}.  Due to potential applications in various
problems of practical interest, synchronization of chaotic elements in
a coupled dynamical system has been an active area of research \cite{laks,piks}.

  Spatially extended systems are suitably modeled by coupled map
lattices (CML).  In comparison to partial differential equations, CMLs
are more suitable for computational studies because of the
discrete nature of time and space while all the
analytical aspects of dynamical systems theory can also be used.  CMLs
were introduced as a simple model for spatio-temporal chaos
\cite{kaneko84}.  They show a variety of phenomena from regular
periodic behavior to very complicated spatio-temporal patterns,
chaos, intermittency, etc \cite{kantsu}.  In CMLs, the dynamical
elements are situated at discrete points in space, 
time is discrete, and the state variable is continuous.
Each spatial unit is coupled to its neighbors.  The selection of
neighbors is determined by the structure of the network.  In most 
studies diffusive coupling (nearest neighbor interaction) is
used.  There are studies on CMLs with various coupling schemes, such as
open network, random network, global coupling , etc \cite{kancml}.
  In most studies a symmetric coupling matrix is employed.

Here we study the synchronization properties of systems formed by a
large number of identical dynamical elements that are connected by
identical symmetrical links.  We derive general conditions  for the
stability of spatially homogeneous solutions of a CML with any
symmetric interaction matrix making use of the spectral properties of
the interaction matrix.  The coupling topology can affect crucially
the synchronizability of the system.

In the next section we describe the properties of the spectrum of the
CML.  We perform a linear stability analysis and give the conditions for
the stability of synchronous solutions and different regimes of
stability.  This is given in Sec.~III.  Results on the global stability
analysis are given in Sec.~IV.  In Sec.~V we provide some
numerical results to elucidate the analytical results with specific
examples.  Here we take a quadratic map for the site dynamics as an
example but the results are valid for any dynamical system.  The results
are even more general in the sense that their validity is not restricted
just to CMLs, but can further be applied almost directly to 
partial differential equations,
coupled ordinary differential equations, etc.  Finally we
provide a discussion on related aspects of the dynamics of CMLs.

\section{CML and its Spectrum}

We consider a coupled map lattice of the form,
\begin{eqnarray}
u(x,n+1) = &&\epsilon \left( {1\over{ n_x}} \sum_{\substack{y \\ x \sim y} } 
f(u(y,n)) 
- f(u(x,n))\right)\nonumber \\ && + f(u(x,n)), \label{eqn1.1}
\end{eqnarray}
where $n_x$ denotes the number of neighbors of $x$. Here, $f:R \rightarrow R$ 
is some differentiable function, often chosen to be 
the quadratic (logistic) map in the literature. $x$ is a spatial variable, its 
domain being 
some finite discrete set $M$. That set carries a neighborhood relationship, 
specifying which $y \in M$ are neighbors of a given $x$ (notation: $x \sim y$). The extreme case is 
the one of a global coupling where all $y$ are neighbors of any $x$. If $M$ has 
the structure of a $k$-dimensional periodic grid, the other extreme case is the 
one of nearest neighbor coupling where only those $y$ are neighbors of $x$ 
that are one step away from $x$ in one of the coordinate directions. In that 
case, each $x$ has $2k$ neighbors. Of course, we also have the trivial case 
where each $x$ is its own neighbor, but has no other neighbors. That case 
of course, represents the absence of coupling.

In the sequel, the only assumption we shall need is that the neighborhood 
relationship  is symmetric, i.e., if $y$ is a neighbor of $x$, then $x$ in turn 
is a 
neighbor of $y$. We also adopt the - completely inessential - convention that $x$ is 
not considered as a neighbor of itself. (Abandoning that convention would 
simply  
amount to a redefinition  of the value of $\epsilon$.)  Finally in order to 
avoid trivial case distinctions, we assume that the neighborhood relationship 
is connected in the sense that for any given $x_1, x_2 \in M$, we find 
$y_1=x_1,y_2,\ldots,y_m=x_2,$ s.t. $y_{j+1} $ is a neighbor of $y_j$ for 
$j=1,2,\ldots,m-1$.  We consider $n=0,1,2,\ldots,$ as the time variable of the 
evolution.

Our subsequent analysis will not depend in conceptual terms on the detailed 
structure of $M$. Of course, the numerical values of the bifurcation parameters 
below will reflect the geometry of $M$.

Our analysis is phrased in general terms and so it is straightforward to extend 
it to the cases:
\begin{itemize}
\renewcommand{\labelitemi}{--}
\item  where  $f$ is vector valued, 
\item where $M$ is a continuous space 
which then has to carry a measure $d\mu$, and the averaged sum needs to be 
replaced by an averaged integral,
\item to weighted neighborhoods i.e., where we 
are given a nonnegative function
$$h:M\times M \rightarrow R^{+}$$ that is symmetric $(h(x,y)=h(y,x) \; \; 
\forall \; x,y 
\in M$) and consider in place of the averaged sum in equation (\ref{eqn1.1})
$$ (1/\sum_y h(x,y)) \sum_y h(x,y) f(u(y,n)), $$ (the situation in (\ref{eqn1.1}) 
corresponds to the choice $h(x,y)= \begin{cases}1 & \text{if} \; \; x,y \; \; 
\mbox{neighbors};\\ 0 & \text{else} \end{cases} $),
\item replacing the last term $f(u(x,n))$ in (\ref{eqn1.1}) by $g(u(x,n))$ for some 
function $g$, 
\item as well as to the case of coupled ordinary differential equations 
in place of difference equations.
\end{itemize}
 As these extensions are rather trivial, we 
refrain from carrying them out.

	The following represents a generalization of the linear stability 
analysis that has been carried out in the literature for some special cases 
such as global coupling \cite{kan90}, nearest neighbor coupling \cite{gade93}, 
 and  random coupling \cite{gade96}.

We shall need the $L^2$-product for functions on $M$:
$$(u,v) := {1\over{\vert M \vert}} \sum_{x\in M} n_x u(x) v(x), $$
where $\vert M \vert$ stands for the number of elements of $M$. We also put 
$||u|| :=(u,u)^{1/2},  (L^2$-norm of $u$). We consider the operator,
$$ \mathcal{L} : L^2 (M) \rightarrow L^2 (M).$$
\begin{equation}
\mathcal{L} v (x) := {1\over {n_x }} \sum_{\substack{y\\x \sim y}} v(y) -v (x). 
\label{eqn1.2}
\end{equation}

  	$\mathcal{L}$ has the following properties:

(i) $\mathcal{L}$ is selfadjoint w.r.t (.,.):
   $$(u,\mathcal{L}v)=(\mathcal{L}u,v)$$ for all $u,v \in L^2(M).$ This follows 
from the symmetry of 
the neighborhood relation.

(ii) $\mathcal{L}$ is nonpositive:
$$(\mathcal{L}v,v) \le 0.$$ This follows from the Cauchy-Schwarz inequality.

(iii) $$\mathcal{L}v=0 \Longleftrightarrow v \equiv  \mbox{constant}.$$

Hence, (i) implies that the eigenvalues of $ \mathcal{L}$ are real. By (ii), they are 
nonpositive; 
we write them as $-\lambda_{k},$ and the eigenvalue equation then is 
$$\mathcal{L} u_k + 
\lambda_k u_k = 0.$$ We order the eigenvalues as $ \lambda_0 \le \lambda_1 \le
\lambda_2  \le \cdots \le \lambda_K$. (This convention deviates from the one used 
in the literature. Our operator $\mathcal{L}$ corresponds to the interaction 
matrix minus 
the identity matrix, and one usually considers the eigenvalues of the former in 
descending order.)

We may then find an orthonormal basis of $L^2 (M),$
$$(u_k)_{k=1,\ldots,K}$$ of eigenvectors of $\mathcal{L}$. 

By (iii) the smallest among the $\lambda_k$ is
$$\lambda_0 = 0,$$ and this is a simple eigenvalue (because we assume that the 
neighborhood relationship is connected), i.e.,
\begin{equation}
\lambda_k > 0 \; \; \mbox{for} \; \; k>0. \label{eqn1.3}
\end{equation}

The numerical values of the bifurcation parameters occurring below will depend 
only (besides on $\epsilon$ and the Lyapunov exponent of $f$) on the eigenvalue 
spectrum of $\mathcal{L}$. This eigenvalue spectrum, of course, reflects the underlying 
geometry of $M$ and of the coupling. Some general considerations may be helpful 
for understanding this point.

In the case of global coupling (including self coupling), we have

$$\lambda_0 = 0 \;  \mbox{(as always)}$$
and
$$\lambda_1 = \lambda_2 = \cdots = \lambda_k = 1, $$
since
$$\mathcal{L}v=-v$$ 
for any $v$ that is orthogonal to the constant map,
i.e., satisfies $$ {1\over{|M|}} \sum_{y \in M}  v(y) =0. $$  If we 
shrink 
the neighborhood size, then the eigenvalues can separate and grow, and in 
particular, the largest one, $\lambda_K$, will become larger the smaller the 
neighborhood size is. In particular, 
$$\lambda_K > 1$$ as there may exist $v \in L^2(M)$ with
$$\sum_{x\in M}   \sum_{\substack{y\\ x \sim y}}  v(x) v(y) < 0$$
(e.g. $M=\{1,2,\ldots,m\}, m $ even, $m>2$, with $\mu$ having neighbors $\mu-1$ 
and 
$\mu+1$, closed periodically, i.e. $m+1 \equiv 1$, $$v(\mu) =\begin{cases} 1,& 
\mu \; \text{even},\\ -1,& \mu \; \text{odd} \end{cases}.$$

Conversely, if the neighborhood interaction matrix of all points   
is the same and kept fixed while we increase the size of $M$, then all 
eigenvalues will decrease. This is a version of Courant's monotonicity theorem 
\cite{courhilb}.  
Thus, from our analysis below, synchronization will require, if possible at 
all, a larger value of the  coupling parameter $\epsilon$.

We also have the following version of Courant's nodal domain theorem 
\cite{nodal}:

\begin{lemma} Consider $M$ as a graph $\Gamma_M$, with an edge between $x$ and $y$ 
precisely if $x$ and $y$ are neighbors. Let $u_k$ be an eigenfunction for the 
eigenvalue $\lambda_k$, with our above ordering, $0=\lambda_0 < \lambda_1 \le 
\lambda_2 \le \cdots \le \lambda_K.$ Delete from the graph $\Gamma_M$ all edges that 
connect points on which the values of $u_k$ have opposite signs. This divides 
$\Gamma_M$ into connected components $\Gamma_1,\ldots,\Gamma_l$. Then $l \le 
k+1.$
\end{lemma}

\section{Linear Stability Analysis}
We now consider a solution $ \bar u (n)$ of the uncoupled equation,
\begin{equation}
\bar u (n+1)= f(\bar u (n)).\label{eqn1.5}
\end{equation}

Clearly, $u(x,n) =\bar u(n)$ then is a solution of (\ref{eqn1.1}). This solution 
is 
spatially homogeneous, or as one says, synchronized. The synchronization 
question then is whether for certain values of the coupling parameter 
$\epsilon$, any solution of  (\ref{eqn1.1}) asymptotically approaches a 
synchronized one. A somewhat weaker question is whether, when we consider a 
perturbation
\begin{equation}
u(x,n) = \bar u (n) + \delta \alpha_k(n) u_k (x), \label{eqn1.6}
\end{equation}
by an eigenmode $u_k$ for some $k \ge 1,$ and small enough 
$\delta, \;\; \alpha_k (n)$ goes to 0 for $n \rightarrow \infty, $ if $u(x,n)$  solves 
(\ref{eqn1.1}). That question can be investigated by linear stability analysis 
and 
we proceed to carry that out. Inserting (\ref{eqn1.6}) into (\ref{eqn1.1}) and 
expanding 
about $\delta=0$ yields
\begin{equation}
\alpha_k(n+1) = (1 -\epsilon \lambda_k) f' (\bar u(n)) \alpha_k (n), 
\label{eqn1.7}
\end{equation}
$f'$ denoting the derivative of $f$. So the sufficient local stability condition
\begin{equation}
\lim_{N \rightarrow \infty} {1\over{N}} \log {{\alpha_k(N)}\over{\alpha_k(0)}} = 
\lim_{N \rightarrow \infty} {1\over{N}} \log \prod_{n=0}^{N-1} 
{{\alpha_k(n+1)}\over{\alpha_k(n)}} < 0 \label{eqn1.8}
\end{equation}
becomes	

\begin{equation}
\log |1 -\epsilon \lambda_k| + \lim_{N \rightarrow \infty} {1\over{N}} 
\sum_{n=0}^{N-1} \log |f'(\bar u(n))| < 0. \label{eqn1.9}
\end{equation}
Here,
$$ \mu_0 = \lim_{N \rightarrow \infty} {1\over{N}} \sum_{n=0}^{N-1} \log 
|f'(\bar u(n))| $$
is the Lyapunov exponent of $f$ and so the stability condition (\ref{eqn1.9}) is 
\begin{equation}
| e^{\mu_o} (1 - \epsilon \lambda_k)| < 1. \label{eqn1.10}
\end{equation}
We may have
\begin{equation}
\mu_o > 0, \label{eqn1.11}
\end{equation}
i.e. temporal instability, but (\ref{eqn1.10}) for all $k \ge 1$; i.e. 
synchronization. We shall now assume (\ref{eqn1.11}) for the remainder of this 
section.
By our ordering convention for the eigenvalues, (\ref{eqn1.10}) holds for all $k 
\ge 1$ if
\begin{equation}
{{1-e^{-\mu_o}} \over {\lambda_1}} < \epsilon < {{1+e^{-\mu_o}} \over 
{\lambda_K}}. \label{eqn1.12}
\end{equation}
In order to satisfy that condition, we need
\begin{equation}
{{\lambda_K}\over{\lambda_1}} < {{e^{\mu_o}+1} \over 
{e^{\mu_o}-1}}.\label{eqn1.13}
\end{equation}
By our above discussion this hold in the globally coupled case because there 
$ \lambda_K = \lambda_1$. By way of contrast if we have nearest neighbor 
coupling, this can only hold if the size of M is not too large. (For a 
1-dimensional chain,  the critical size is 
5, with a large value of $\epsilon.$ If we have second nearest neighbor 
coupling, the critical 
size of a one dimensional chain is 9.)

Let us now assume that (\ref{eqn1.13}) holds. We then predict the following 
behavior of the coupled system as $\epsilon$ increases.

For very small values of $\epsilon > 0$, as we assume (\ref{eqn1.11})
$$e^{\mu_o} (1 - \epsilon \lambda_k) > 1, $$ and so, all spatial modes $u_k, 
k\ge 1,$ are unstable, and no synchronization occurs. If we are in the globally 
coupled case, then there exists a single critical value $\epsilon_c$ such that 
$$e^{\mu_o} (1 - \epsilon_c \lambda_k) = 1 $$ for all $k=1,2,\ldots,K$. For 
$\epsilon > \epsilon_c,$ the dynamics become synchronized. For $\epsilon$ 
slightly smaller than $\epsilon_c,$ one observes intermittent behavior, 
clustering, etc \cite{kaneko89}. 

Let us now consider the more interesting case where the coupling is not global 
so 
that not all the $\lambda_k$ are equal; in particular
$$ \lambda_1 < \lambda_K.$$ We then let $\epsilon_k$ be the solution of 
$$e^{\mu_o} (1 - \epsilon_k \lambda_k) = 1 $$
The smallest among these values is $\epsilon_K$, the largest $\epsilon_1$.
If now, for $k_1 < k_2,$
$$\epsilon_{k_2} < \epsilon < \epsilon_{k_1}$$
then the modes $u_{k_2}, u_{k_{2}+1},\ldots,u_K$ are stable, while the modes
$u_1, u_2,\ldots,u_{k_1}$ are unstable. Because of Lemma 1, we see that
desynchronization can lead to utmost $k_2 +1 $ subdomains on which the
dynamics is either advanced or retarded.

In particular, if $\epsilon$ increases, first the highest modes, i.e., the ones 
with most spatial oscillations, become stabilized, and the mode $u_1$ becomes 
stabilized the last. So if $\epsilon_2 < \epsilon < \epsilon_1$, then any 
desynchronized state consists of two subdomains.

We then let $\bar{\epsilon_k}$ be the solution of
$$e^{\mu_o} (\bar{\epsilon_k} \lambda_k - 1) = 1 $$
Again,
$$\bar{\epsilon_k} \le \bar \epsilon_{k-1}.$$
Because of (\ref{eqn1.12}),
$$\epsilon_1 < \bar \epsilon_K.$$
If $$\epsilon_1 < \epsilon < \bar \epsilon_K,$$
then all modes $u_k, k=1,2,\ldots,K,$ are stable, and the dynamics synchronizes.

If $\epsilon$ increases beyond $\bar \epsilon_K$, then the highest frequency 
mode $u_K$ becomes unstable and we predict spatial oscillations of high 
frequency of a solution of the dynamics . If $\epsilon$ increases further then 
more and more spatial modes become destabilized.

\section{Global Stability Analysis}
The basis of the preceding analysis was a linear expansion about a synchronized 
state $\bar u(n).$ Therefore, that analysis is valid only for small 
perturbations about such a state. 	In this section, we want to derive a 
criterion that guarantees synchronization for arbitrary starting values $u(x,0)$ 
of a solution of (\ref{eqn1.1}).

From general principles of functional analysis (see \cite{yosida}), there exists an 
operator,
$$ \Lambda : L^2(M) \rightarrow L^2(M)$$ with
\begin{equation}
- (u, \mathcal{L} v) = (\Lambda u, \Lambda v), \forall u,v \in L^2(M). 
\label{eqn2.1}
\end{equation}
This follows from the self adjointness of $\mathcal{L}$. It is not difficult to 
write a 
$\Lambda $ down explicitly, but our more abstract approach provides the 
advantage of a less cumbersome notation.

$\Lambda$ is nonnegative in the sense that
\begin{equation}
(\Lambda u, \Lambda u) \ge 0, \forall u\in L^2(M), \label{eqn2.2}
\end{equation}
and we even have
\begin{equation}
\Lambda u =0 \Longleftrightarrow u \equiv \mbox{constant} \label{eqn2.3}
\end{equation}
(This follows from the nonpositivity properties of $\mathcal{L}$).

Moreover $\Lambda$ commutes with $\mathcal{L}$, i.e.,
\begin{equation}
\Lambda \mathcal{L} = \mathcal{L} \Lambda, \label{eqn2.4}
\end{equation}
and so, we may assume that the $u_k$ are also eigenfunctions of $\Lambda$.

Therefore a natural ansatz for a Lyapunov function for the dynamics 
(\ref{eqn1.1}) 
is
\begin{equation}
\Phi(n) := (\Lambda u(.,n),\Lambda u(.,n)),\label{eqn2.5}
\end{equation}
and it remains to derive conditions under which
\begin{equation}
\Phi(n) \rightarrow 0, \mbox{for}\;\; n\rightarrow \infty. \label{eqn2.6}
\end{equation}
We have
\begin{eqnarray}
 \Phi(n+1) & = & ( \Lambda u (., n+1), \Lambda u (., n+1)) \nonumber \\
   & = & (\Lambda u(.,n+1),\Lambda(\epsilon L f(.,n) + (1-\epsilon) f (.,n))) \nonumber 
\end{eqnarray} by 
(\ref{eqn1.1}).

Since the $u_k$ are an orthogonal basis of $L^2(M)$, we may write
$$f(u(x,n)) = \sum_{k=0}^K \beta_k (n) u_k (x),$$ with 
$\beta_k(n)=(f(u(.,n)),u_k)$.
Inserting this into the last equality, we get
\begin{equation}
\Phi(n+1) = \left( \Lambda u (., n+1), \Lambda   \sum_{k=0}^K (1-\epsilon 
\lambda_k) \beta_k(n) u_k \right). \label{eqn2.7}
\end{equation}
The important observation now is that in the last sum, we can discard the 
summand $k=0$, because $u_0$ is constant, and so $$\Lambda u_0 = 0.$$
Moreover, we observed above that, since $\Lambda$ commutes with $\mathcal{L}$, we may 
assume,
$$(\Lambda u_k , \Lambda u_l ) = 0, \mbox{for} \;\; k \neq l,$$ and so
$$ || \Lambda f (u (.,n))||^2 = \sum_{k=0}^K \beta_k^2 (n) ||\Lambda u_k 
||^2.$$Using these observations and the Cauchy-Schwarz inequality in 
(\ref{eqn2.6}), we may
estimate
\begin{equation}
\Phi(n+1) \le {1\over {2}} ||\Lambda u(.,n+1)||^2 +{1\over {2}}(1 - \epsilon 
\lambda_1)^2 || \Lambda f (u(.,n))||^2, \label{eqn2.8}
\end{equation}
assuming $| 1 -\epsilon \lambda_K| \le 1 - \epsilon \lambda_1, $ i.e.,
\begin{equation}
\epsilon \le {2 \over {\lambda_1 + \lambda_K}}.\label{eqn2.9}
\end{equation}
If we now use the coarse estimate
\begin{equation}
|| \Lambda f (u(.,n))|| \le \sup |f'| ||\Lambda u (., n)||, \label{eqn2.10}
\end{equation}
we obtain from (\ref{eqn2.8})
\begin{equation}
\Phi(n+1) \le (1 -\epsilon \lambda_1)^2 \sup |f'|^2 \Phi(n).\label{eqn2.11}
\end{equation}

We conclude 

\begin{theorem}
  The coupled dynamical system (\ref{eqn1.1}) asymptotically 
synchronizes if $\epsilon$ satisfies (\ref{eqn2.9}) and
\begin{equation}
(1 -\epsilon \lambda_1) \sup |f'| < 1.\label{eqn2.12}
\end{equation}
\end{theorem}
{\bf Remark:} If (\ref{eqn2.9}) does not hold, (\ref{eqn2.12}) needs to be replaced 
by,
\begin{equation}
(\epsilon \lambda_K -1 ) \sup |f'| < 1.\label{eqn2.13}
\end{equation}

In the special case of global coupling, the synchronization condition becomes,
\begin{equation}
(1 - \epsilon  ) \sup |f'| < 1,  (0 \le \epsilon < 1) \label{eqn2.14}
\end{equation}
The reason why we have $\sup|f'|$ in (\ref{eqn2.12}), in place of $ e^{\mu_0}, 
\mu_0$ being the Lyapunov exponent of $f$, as in section 3, is that here we do 
not linearize about a spatially homogeneous solution. Our global approach rather 
requires to consider any solution $u(x,n)$ of (\ref{eqn1.1}). This means, 
however, 
that our condition (\ref{eqn2.12}), while sufficient, need not be necessary for 
synchronization.

\section{Numerical Results}

In this section we demonstrate our results with different coupling
schemes or network topology.  For our numerical study we took the quadratic
map for the site dynamics.  The quadratic map is a widely studied chaotic
map, given by $f(x)= 1 - a x^2$ \cite{coleck}.  Here $a$ is a parameter and varying
its value the single map shows a variety of dynamical phenomena.  It
becomes chaotic when $a \approx 1.4011$, going through a period
doubling bifurcation sequence.  At $a=2$ the map is maximally chaotic,
with a Lyapunov exponent $\mu_0 = \log (2)$.

\begin{figure}
\includegraphics[width=7cm]{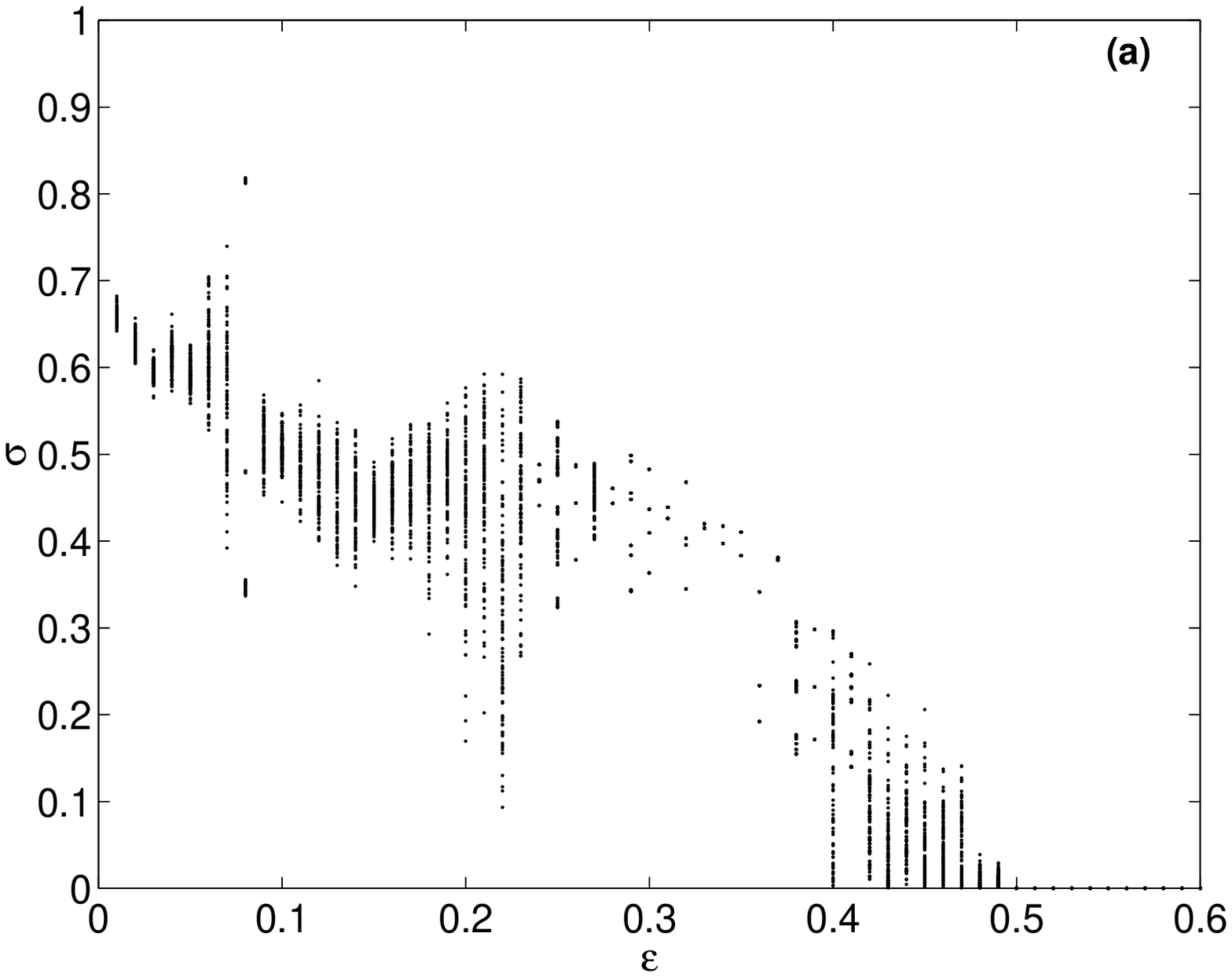}
\includegraphics[width=7cm]{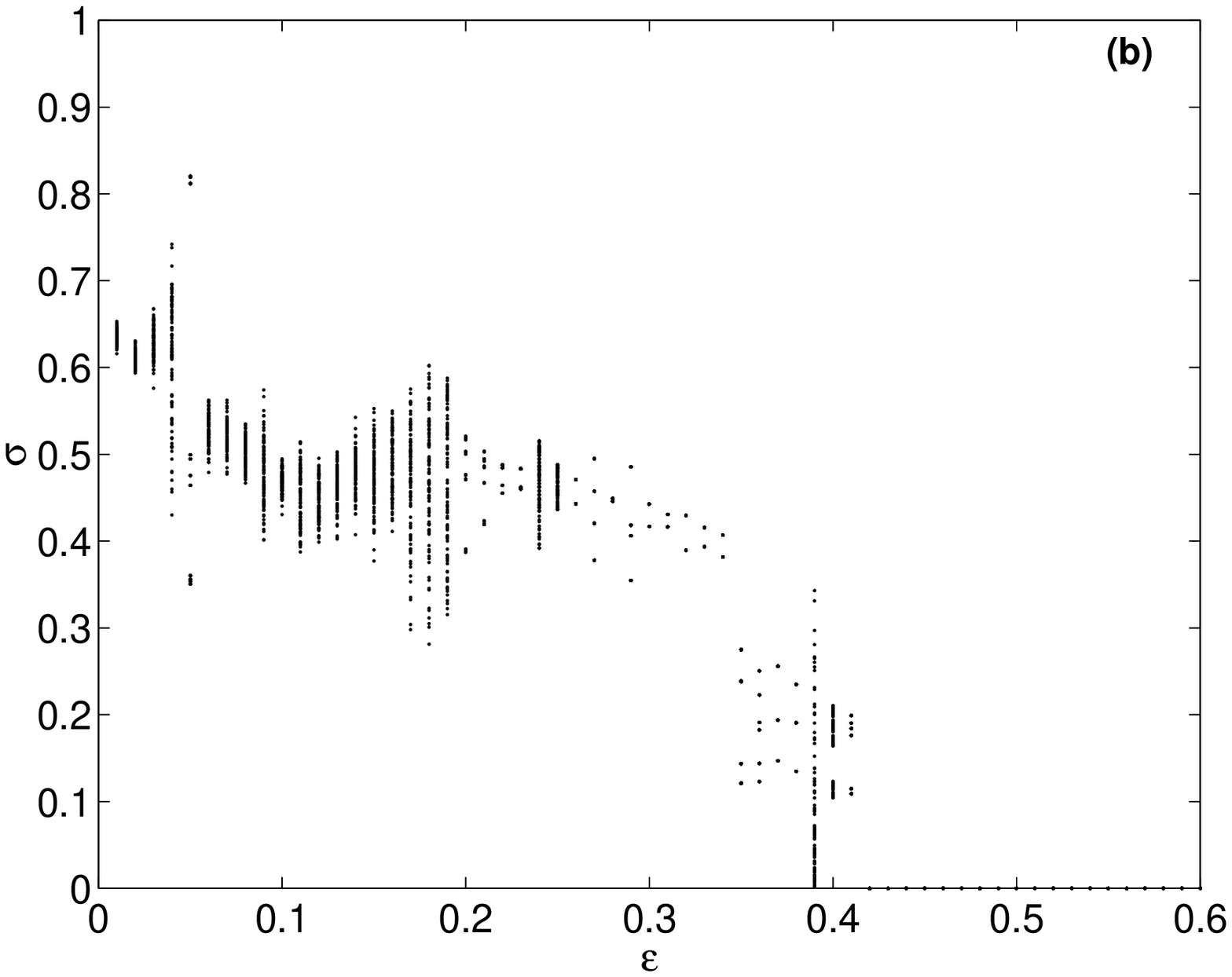}
\caption{Fluctuation of the mean, $\sigma (n)$ is shown as a function of the 
coupling strength, $\epsilon$. At each value of $\epsilon, \;$ 200 
final iterates of 
$\sigma(n)$ are plotted. Here $m=1000$ and the coupling is global. In (a) 
$a=2.0$, and in (b) $a=1.9$ .}
\label{fig1}
\end{figure}

\subsection{Global coupling}

In the case of global coupling, we have $ \lambda_0 = 0$ and
$\lambda_1 = \lambda_2 = \cdots  = \lambda_{m-1} = 1$.  (The self coupling
term is also included here.)  This case has been studied in various
contexts.  When $ \exp (\mu_0) (1 - \epsilon) < 1$, the spatially
homogeneous solution is stable, as shown in \cite{kan90}.  
For the quadratic map with $a=2$, it
becomes stable when $\epsilon > 0.5$.  Just below this value the
system shows spatio temporal intermittency, clustering phenomena, etc \cite{kantsu}.
In Fig.~\ref{fig1}(a), we display $\sigma (n) $, the fluctuation of the state variable
from the mean, defined by $\sigma^2 (n) = \frac{1} {m} \sum_{i=1}^{m} (x_i
(n)- \bar x (n))^2, (\bar x (n) $ is the average of all $x_i (n)$), 
 for different values of $\epsilon$, for the case with $a=2.0$.
It can be seen that when $\epsilon > \epsilon_c = 0.5$, the value of $\sigma$
becomes zero (within the numerical accuracy) indicating that the
system is synchronized.  Though the linear stability does not guarantee
the synchronization from arbitrary initial conditions, in this case it
happens.  We started with random initial conditions for the individual
sites, and after a few iterations the system synchronizes, indicating
the stability of the spatially homogeneous solutions in these
parameter regimes.  For $a=1.9$, the system synchronizes for a smaller
value of $\epsilon$, since the Lyapunov exponent at that parameter
value is 0.5490.  Here the critical value is $\epsilon_c =0.4225$.
Figure~\ref{fig1}(b) gives details of this case.  We took $m=1000$ for
our simulations.

\subsection{Nearest Neighbor Coupling}

Here the eigenvalues   are given by $ \lambda_0 =
0$ and $\lambda_i = 1 - \cos (\frac{2 \pi i}{m}), i=1,2,3,\ldots,m-1$.
The first nonzero eigenvalue is
$$\lambda_1=1 - \cos (\frac{2 \pi }{m}),$$ 
and the largest eigenvalue is
$$\lambda_K~=~\begin{cases} 2 &
\mbox{for even} \; m \\1 + \cos (\frac{\pi}{m}) & \mbox{for odd} \; m
\end{cases}. $$

	Using this one can calculate the maximum value of $m$ at which
the spatially homogeneous solution can be stable using the condition
for linear stability.  It will occur when
$$\frac{\lambda_K}{\lambda_1} < \frac{\exp(\mu_0) +1}{\exp (\mu_0)
-1}$$ and the value of $\epsilon$ lying between
$$\frac{1-\exp(-\mu_0)}{\lambda_1} < \epsilon < \frac{ 1 +
\exp(-\mu_0)}{\lambda_K}.$$

\begin{figure}
\includegraphics[width=7cm]{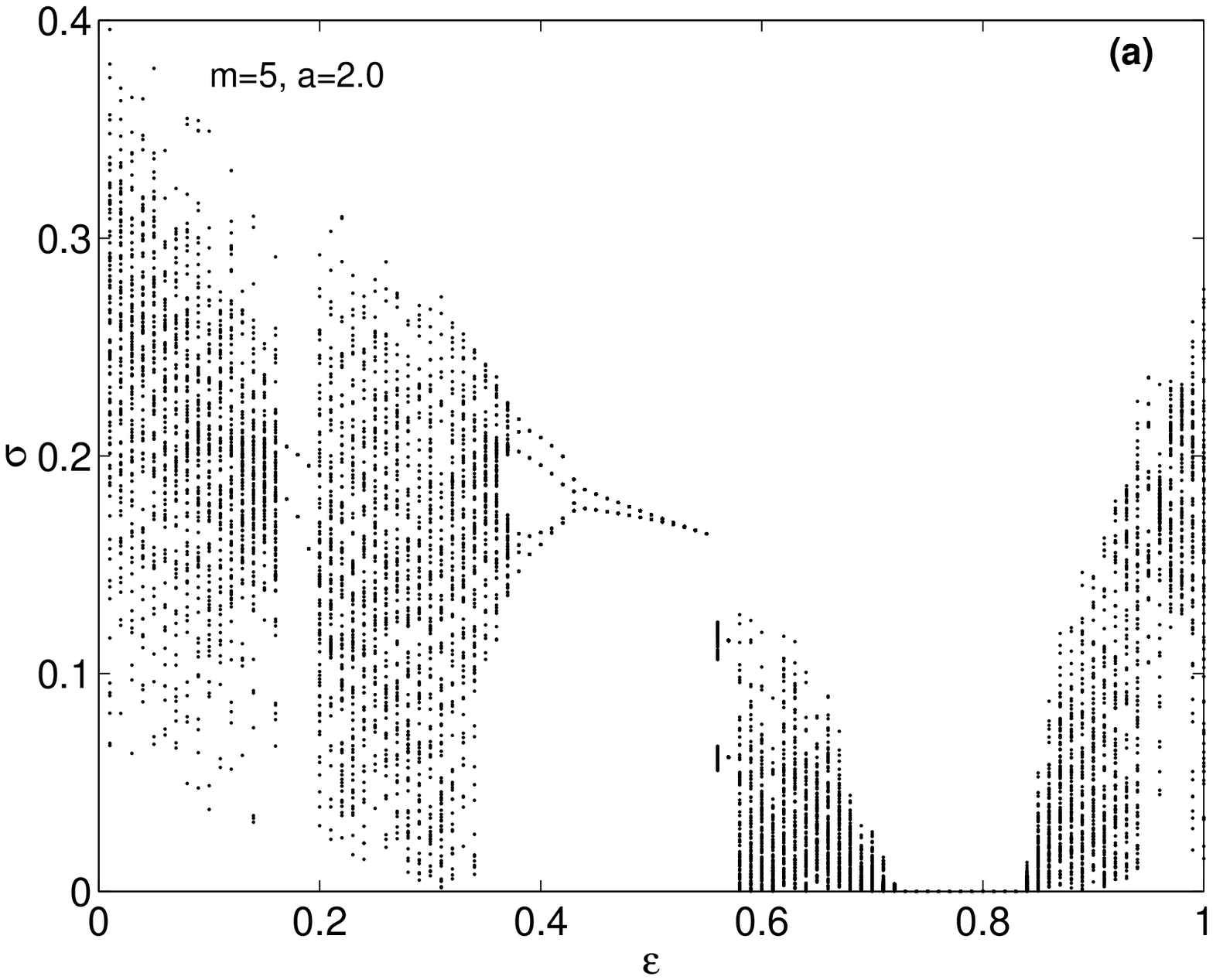}
\includegraphics[width=7cm]{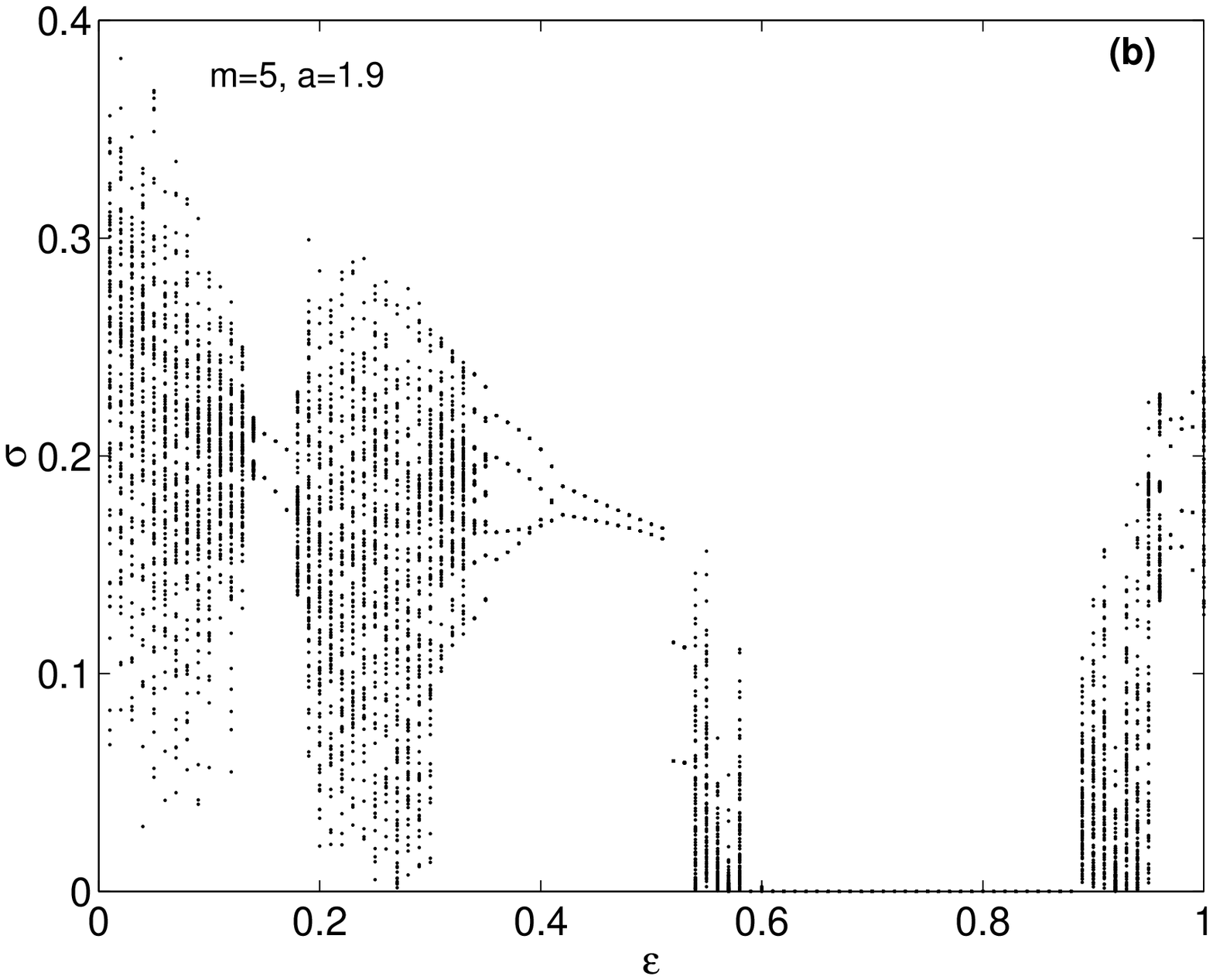}
\caption{Same as in Fig.~\ref{fig1} with nearest neighbor coupling, for $m=5$
, in 
(a) $a=2.0$, and in (b) $a=1.9$.}
\label{fig2}
\end{figure}

\begin{figure}
\includegraphics[width=7cm]{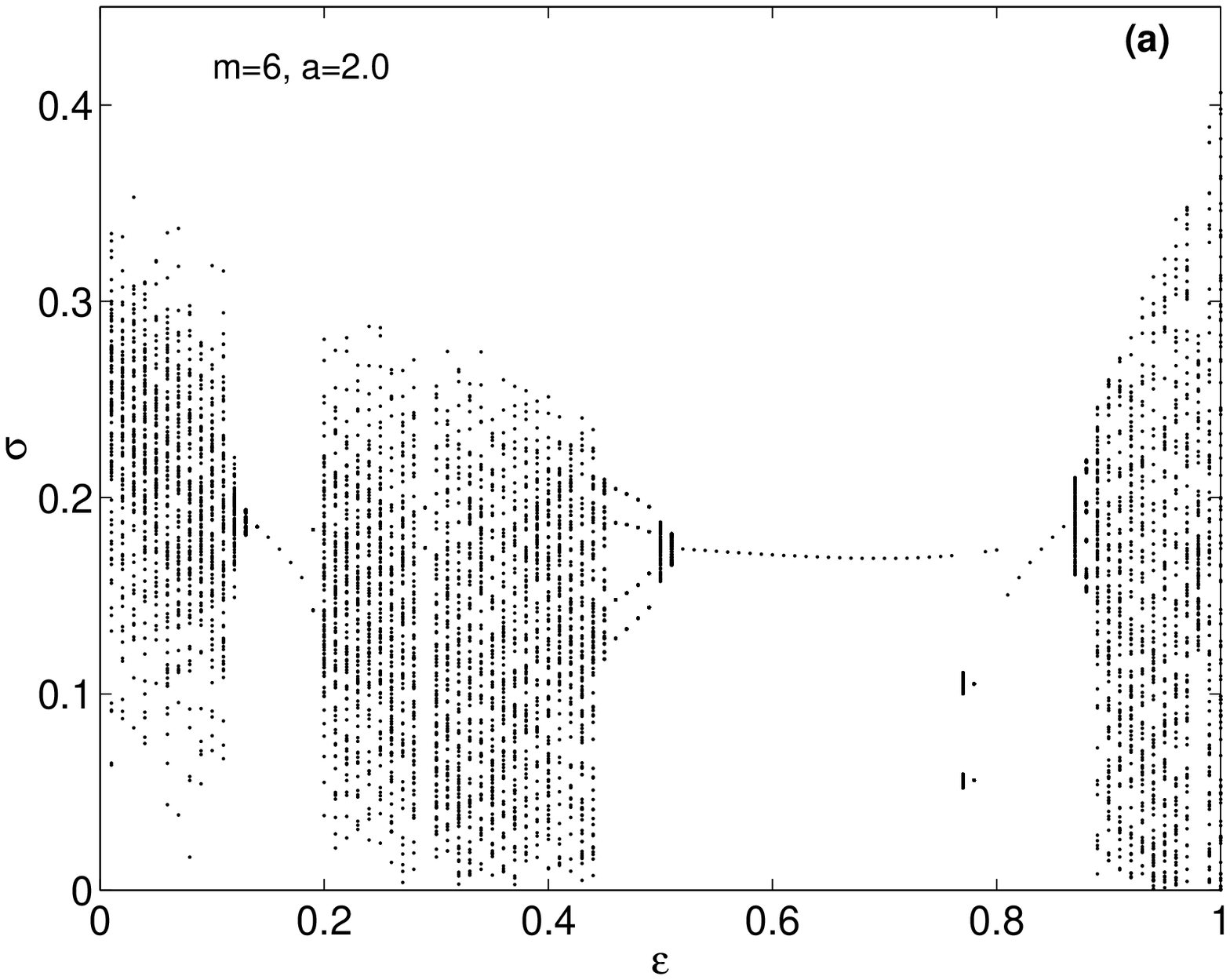}
\includegraphics[width=7cm]{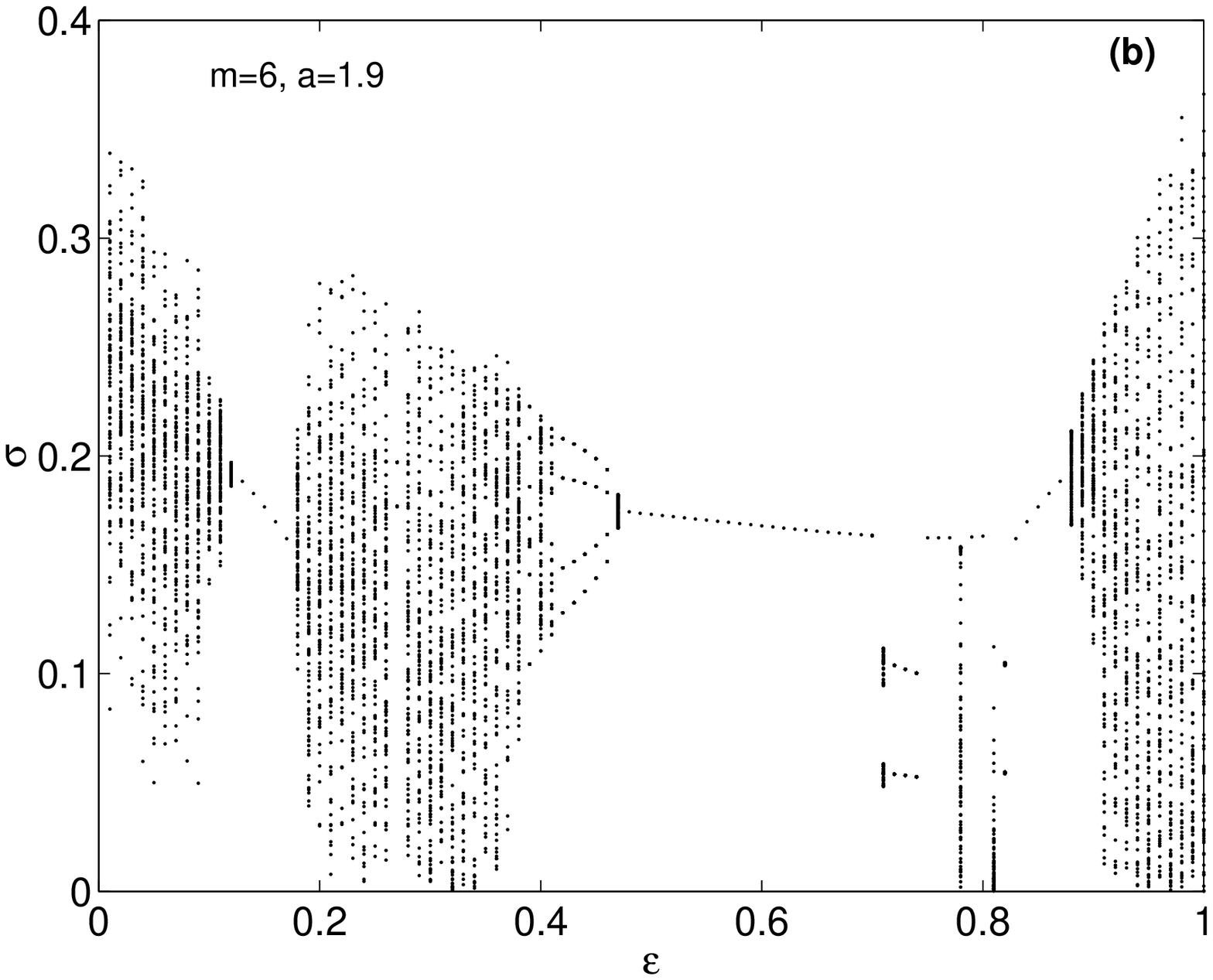}
\caption{Same as in Fig.~\ref{fig2} with nearest neighbor coupling, for $m=6$
.}
\label{fig3}
\end{figure}

	For a CML with a fully chaotic quadratic map the maximum value of the 
system
size which can sustain a stable synchronous solution is $m = 5 $,
when $\epsilon$ is between 0.7236 and 0.8292.  In the case of $m=6$, the
first mode becomes stable at $\epsilon=1$, but the last mode becomes
unstable for a value of $\epsilon $ above 0.75.  Hence there is no
synchronization.  The second mode is stable when $\epsilon$ is 
between 0.333 and 1.  In
Figs.~\ref{fig2}--\ref{fig3} we give the plot of the fluctuation of
the mean field for different values of $\epsilon$, for $m=5$ and $m=6$,
when $a=2.0$ (a) and $a=1.9$ (b).  When $m=6$, between $\epsilon= 0.33$ and 0.75 only one
mode is unstable.  From the spectrum one can see that the largest value of $m$ 
for which
only the first mode is unstable is  $m=9$ when $0.72 < \epsilon <
0.75$. For higher values of $m$ more than one mode will be unstable for any 
value of $\epsilon$. So there cannot be synchronization in large systems with 
nearest neighbor coupling.

\subsection{Intermediate Range Coupling}

If we consider $k$ nearest neighbors  (there will be
$2k$ neighbors for each site) the eigenvalues are given by $\lambda_0 =0$
and $$\lambda_i = 1 - \frac{1}{2k} \sum_{j=1}^{k} \cos (\frac {2 \pi i
j}{m}), i=1,2,\ldots,m-1.  $$

\begin{figure}
\includegraphics[width=7cm]{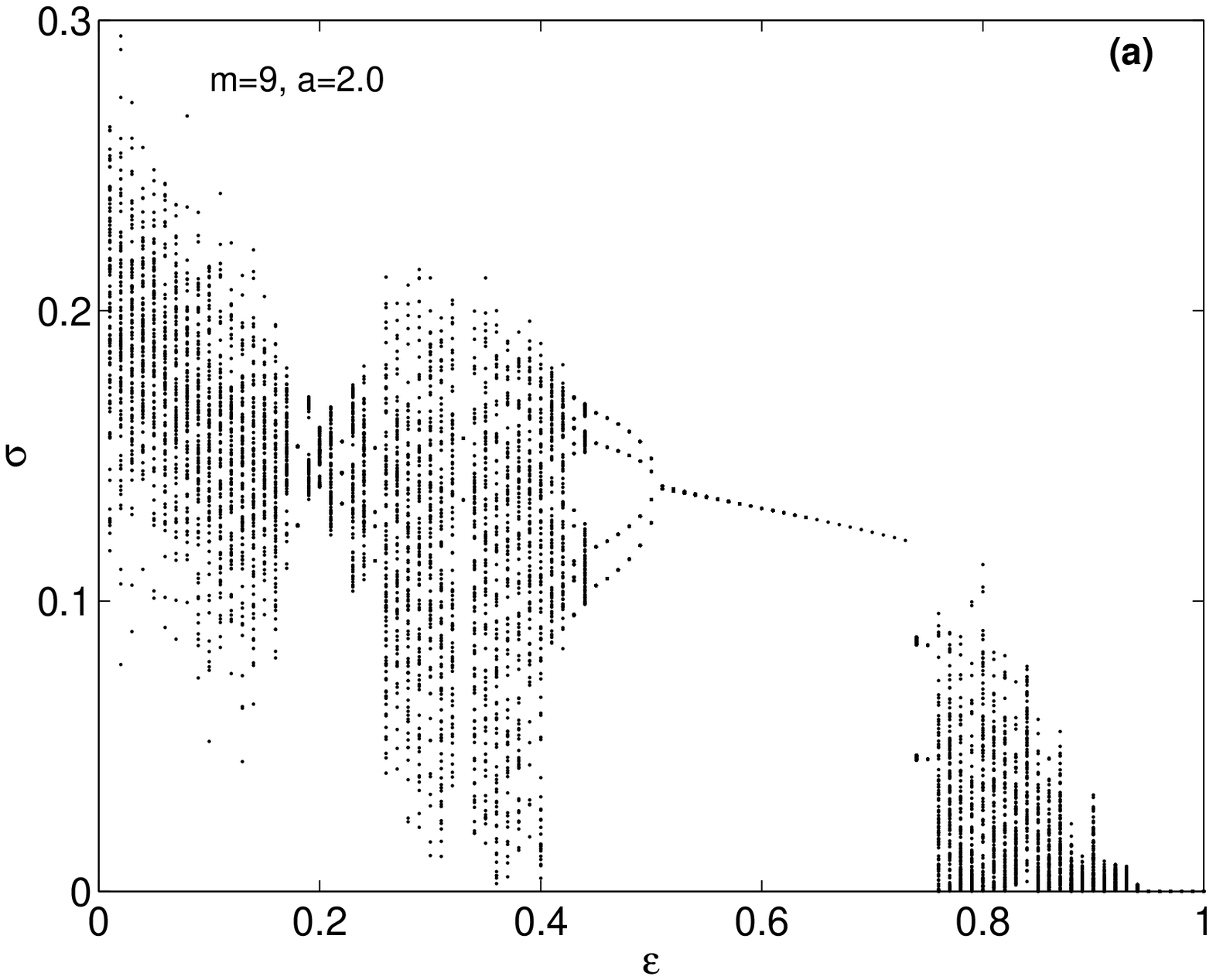}
\includegraphics[width=7cm]{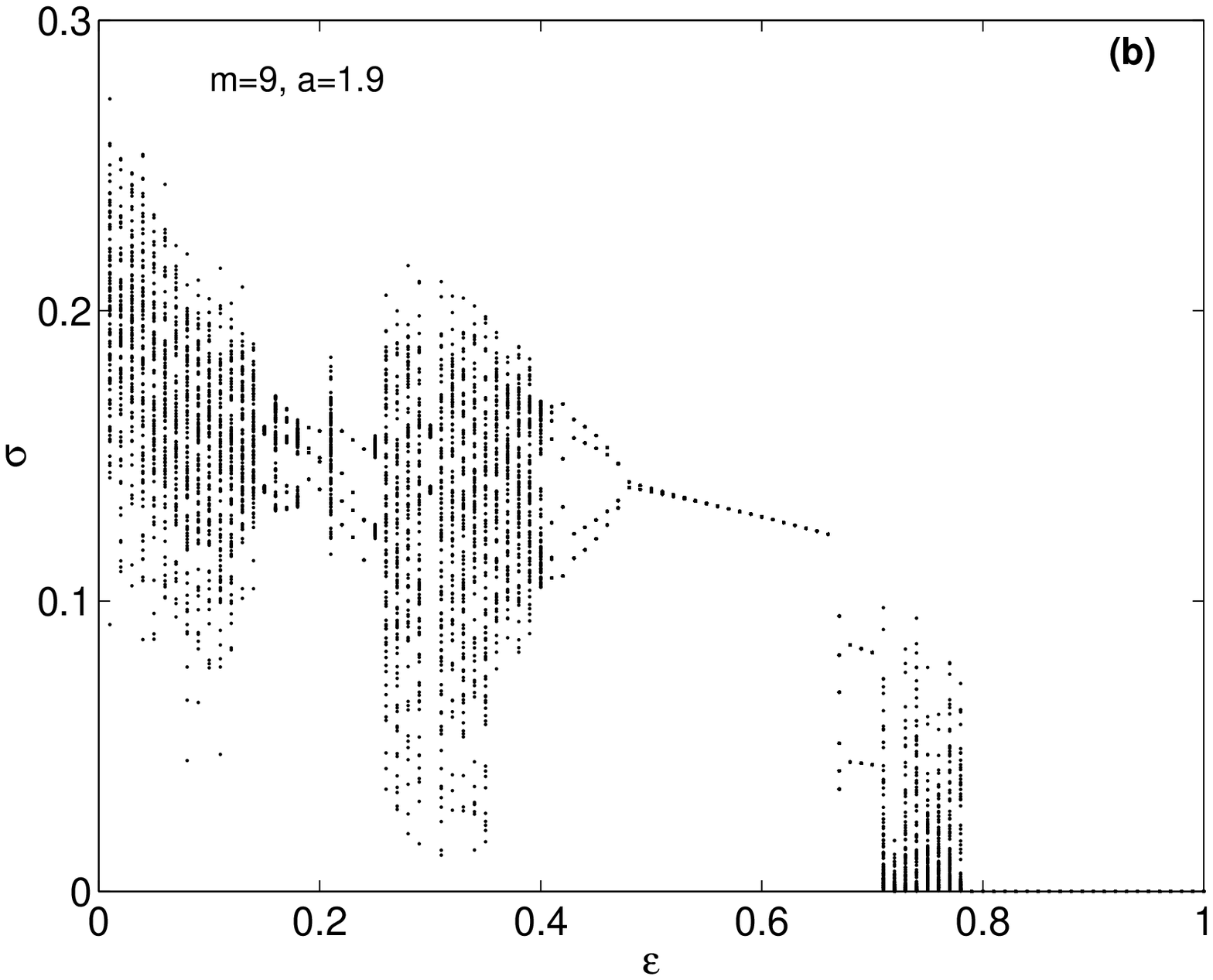}
\caption{Same as in Fig.~\ref{fig2} with two nearest neighbors coupling, for 
$m=9$. }
\label{fig4}
\end{figure}

Let us consider the case of two nearest neighbors ($k=2$).  As in the
case of NN coupling one can find the maximum value of $m$ at which the
CML can sustain stable synchronous chaotic oscillations.  For k=2, it is $m=9$ 
with $0.33 < \epsilon < 1$.  The maximum value of
$m$ at which the second largest mode also becomes unstable is $m=18$.
Figure~\ref{fig4} gives the plot for $m=9$ 
for $a=2.0$ (a) and $a=1.9$ (b).  For 3 nearest neighbors
$(k=3)$, it is at $m=12,$ and for $ k=4, m=15$.

\begin{figure}
\begin{center}
\includegraphics[width=7cm]{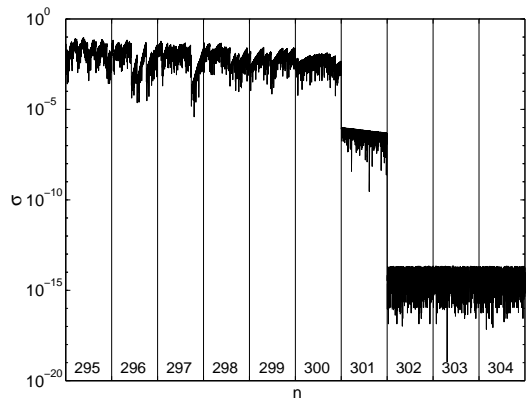}
\caption{Here $\sigma(n)$ is plotted for different values of $k$ 
(denoted inside the grid) with intermediate range coupling, for  $m=1000, 
a=2.0,$ and $ \epsilon = 1.0$. Between two grid lines  1000 iterates of $\sigma(n)$ are
 plotted.}
\label{fig5}
\end{center}
\end{figure}

One can see that for $k/m > 0.301  $ the system
synchronizes when the coupling is strong, i.e., $\epsilon=1$.  In
Fig.~\ref{fig5},  $\sigma (n)$ is shown as a function
of $k$ for $m=1000$ near the synchronization transition region. In each grid
 corresponding to a $k$ value, 1000 final iterates of $\sigma (n)$ are plotted
 after discarding initial transients. The system shows 
synchronization when $k=301$.  As the system size increases we need a higher number 
of neighbors for synchronization. For a fixed number of neighbors the 
behavior is like that of the NN case; there is no synchronization when the 
system size increases.

\subsection{Random Coupling}

Now we consider a case where there is coupling between random
sites.  For every site we randomly select $k$ other distinct
sites and connect them with each other under the constraint that  self and multiple 
coupling is prohibited.  The average degree of a node in such a graph obtained 
is $2k$.  
 For the quadratic map with $a=2.0$ and $\epsilon = 1$, 
 the system synchronizes for  large $m$, if  $k>8$,
in contrast to the unsymmetric case where it does so for $k>4$ 
\cite{gade96}.

\begin{figure}
\includegraphics[width=7cm]{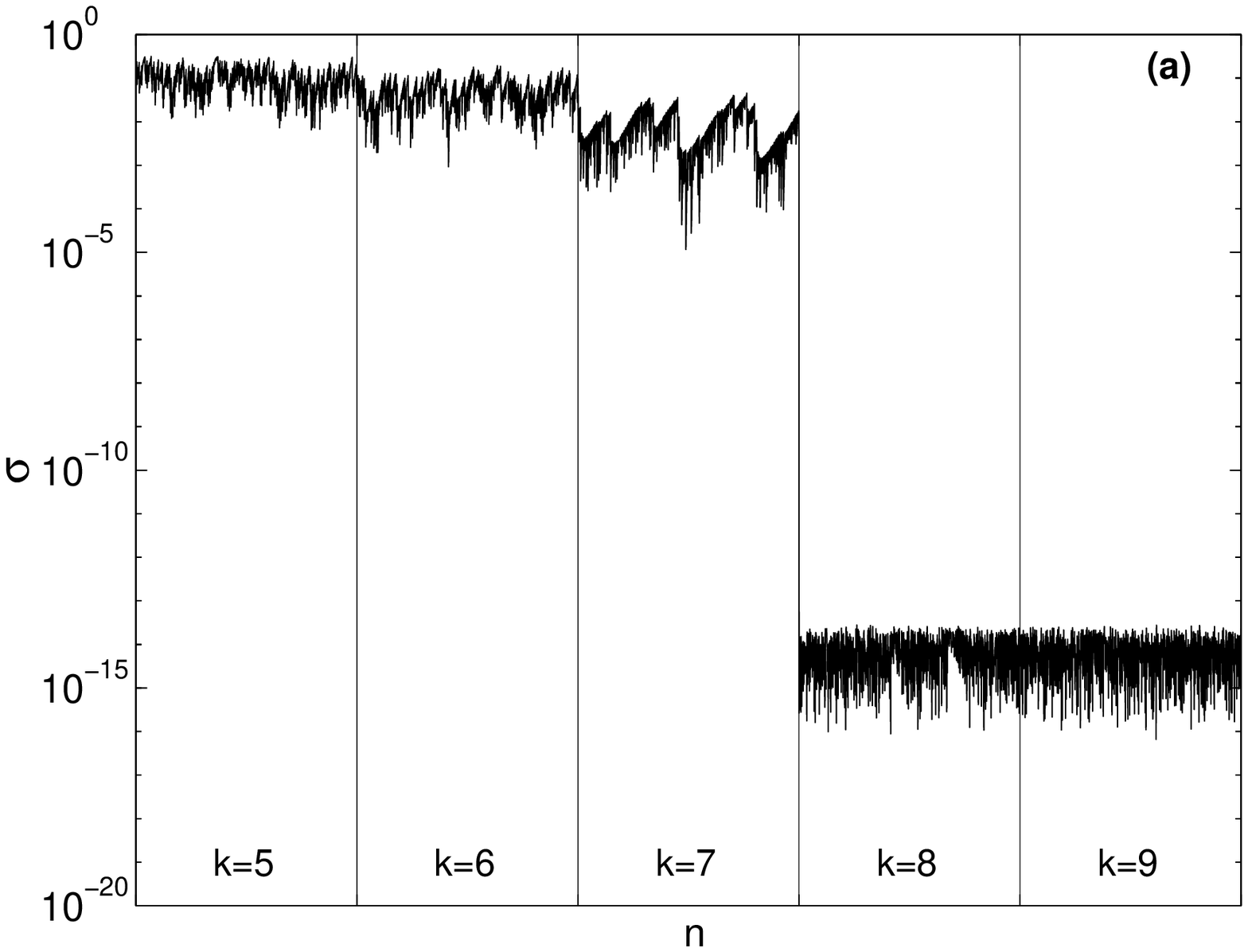}
\includegraphics[width=7cm]{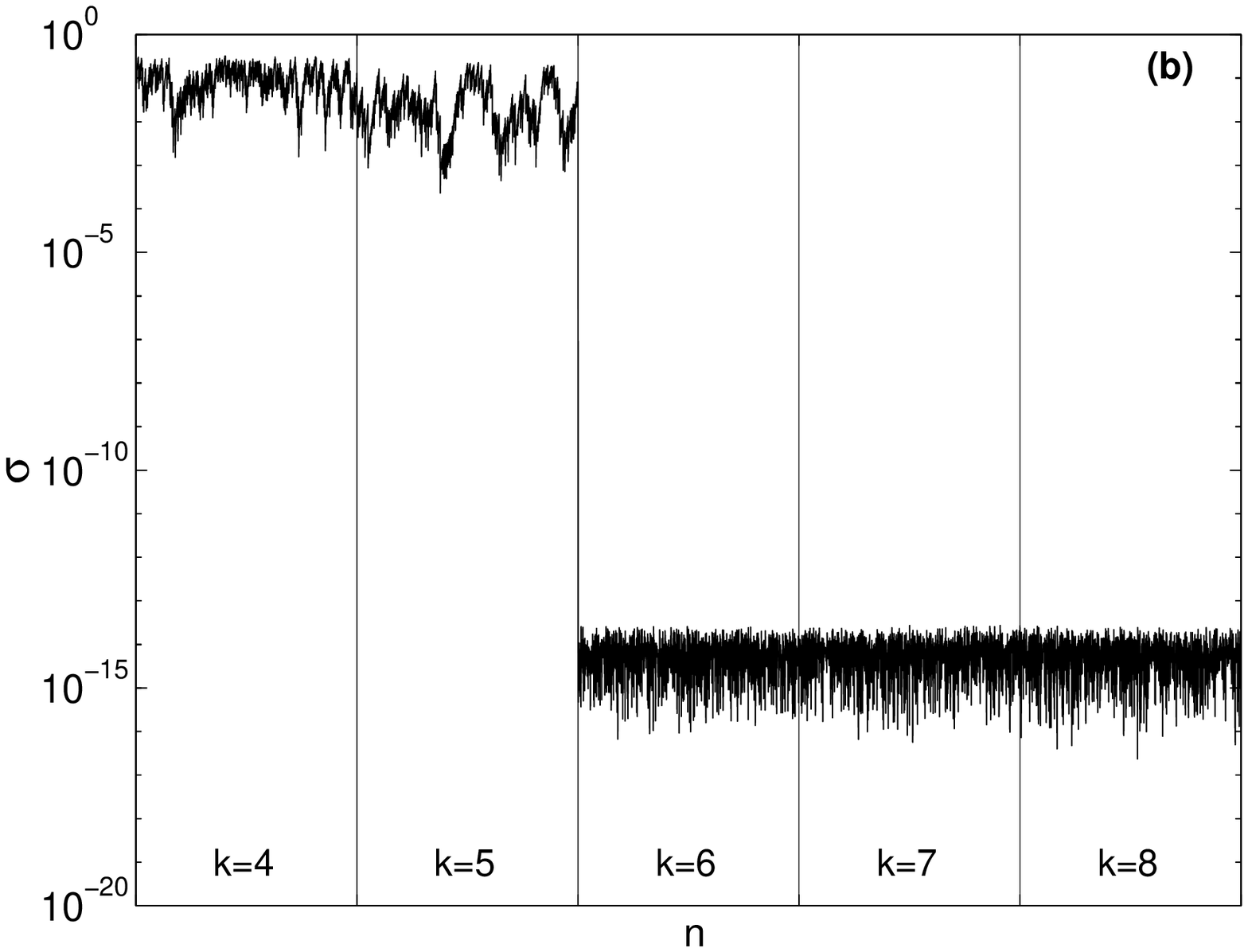}
\caption{Same as in Fig.~\ref{fig5} with random coupling for different $k$ 
values, for $m=1000, $ and $\epsilon= 1.0$. In (a) $a=2.0$ and in b) $a=1.9$.}
\label{fig6}
\end{figure}

   We plot the fluctuation of
the mean field, $\sigma$, for different values of $k$ for $m=1000$, 
$\epsilon =1.0$,  $a=2.0$ (Fig.~\ref{fig6}(a)), and $a=1.9$
 (Fig.~\ref{fig6}(b)).   It can 
be
seen that the system synchronizes when the average degree of a vertex
is 8 or more, for the completely chaotic quadratic map.  This is independent
of the system size $m$.  From random matrix theory one can see that
the value of $\lambda_1$ depends only on $k$ \cite{bollo}.  For smaller $m$,
synchronization can occur below $k=8$ because of the finite system
size effects.  So unlike in the case of nearest neighbor or
intermediate range interactions, in the case of random coupling, one
can have chaotic synchronization for any arbitrarily large value of
$m$, if the number of neighbors ($k$) is larger than some threshold 
determined by the value of the maximal Lyapunov exponent of the chaotic map.

\subsection{Small-world Networks}

Small-world (SW) networks have an intermediate connectivity between regular
and random networks.  They are characterized by a very small mean path length as 
in
random networks while at the same time having a high clustering coefficient as in
regular networks.  SW coupling is done as in the  Watts and Strogatz
algorithm \cite{ws98}.  We start with a lattice of $m$ vertices each
connected to its $k$ neighbors.  With a probability $p$ we reconnect
each edge to a vertex chosen uniformly at random over the entire
lattice.  Duplicate edges are avoided.  It has been shown in
\cite{ws98} that even for a very small random rewiring probability $p$
there is a  transition to the small-world regime.

\begin{figure}
\begin{center}
\includegraphics[width=7cm]{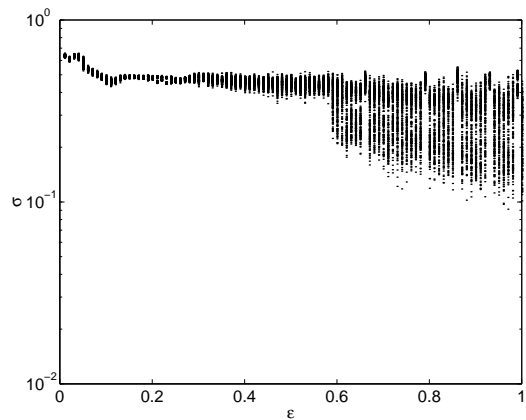}
\caption{Same as in Fig.~\ref{fig1} with small-world coupling, for $m=1000, k=
10$, $a=2.0$, and $p=0.1$.}
\label{fig7}
\end{center}
\end{figure}

\begin{figure}
\begin{center}
\includegraphics[width=7cm]{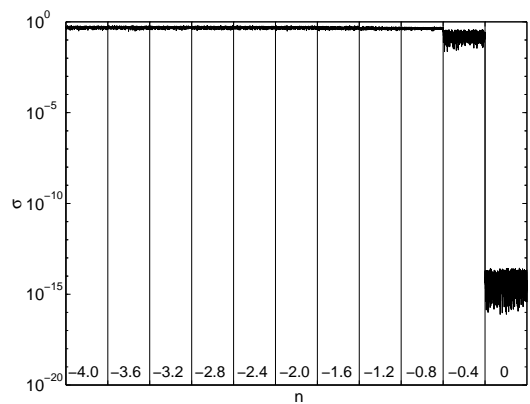}
\caption{$\sigma(n)$ for different values of $p$ with small-world coupling, for 
$m=1000, k=10$, $a=2.0$ and $\epsilon=1$. Between two grid lines, corresponding 
to a $p$ value, 1000 iterates of $\sigma(n)$ are plotted  and the corresponding  
$\log_{10} (p) $ is denoted at the bottom.}
\label{fig8}
\end{center}
\end{figure}

    Here we took $p=0.1$ at which there is small world effect on the
structural properties of the graph.  Figure~\ref{fig7} gives the
fluctuation $\sigma (n)$ for different values of $\epsilon$, 
$k=10$ and $m=1000$.  One can see that there is no synchronization at
this value of $p$, $\lambda_1 \approx  0.08$ . When $p=0.8$, there is 
synchronization for
$\epsilon = 1.0$.  At this value the number of random connections per
vertex reaches the value needed for synchronization. In Fig.~\ref{fig8}, 
$\sigma(n)$ is plotted for $p$ values from 0.0001 to 1, and $\epsilon=1.0$. Between
two grid lines 1000 iterates of $\sigma(n)$ are plotted  and the corresponding
$\log_{10}(p)$ values are also denoted. From this figure we can easily see that there is no
synchronization for smaller $p$  values.

\subsection{Scale-free Networks}

\begin{figure}
\begin{center}
\includegraphics[width=7cm]{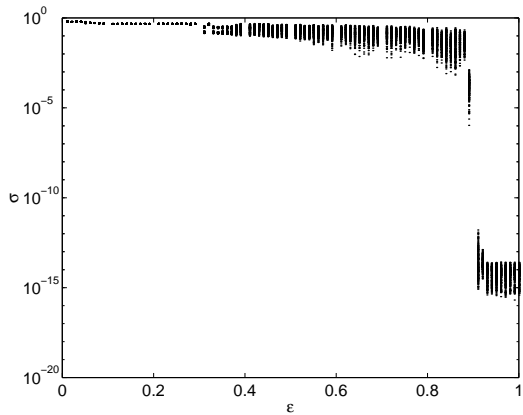}
\caption{Same as in Fig.~\ref{fig1} with  scale free coupling, for $m=1000, k=
6$ and $a=1.9$.}
\label{fig9}
\end{center}
\end{figure}

\begin{figure}
\begin{center}
\includegraphics[width=7cm]{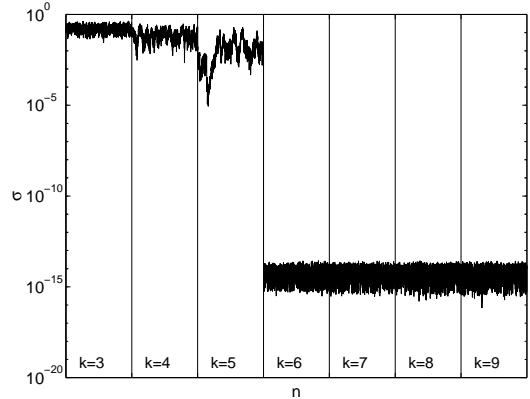}
\caption{Same as in Fig.~\ref{fig5} with scale free coupling, for $m=1000, a
=1.9$ 
and $\epsilon=1.0$.}
\label{fig10}
\end{center}
\end{figure}

Another widely studied class of networks are the scale free networks,
where the degree distribution obeys a power law which is observed in
many real networks.  We studied the synchronization of a scale free
network constructed by the  Barabasi-Albert algorithm \cite{ba99}.  We
start with  $k_0$ vertices and at every time a new node is
introduced.  The new node is connected to $k$ already existing nodes
and they are selected with a probability proportional to the degree of
that node.  The process is continued for a long time and then the
degree distribution is described by the power law, $P(k) \sim 
k^{-\gamma}$, where $\gamma =3 $.  It is independent of $k_0$.  For
this study we took $k_0 = k$ and a network of size $m=1000$. 
Figure~\ref{fig9} shows the $\sigma$ versus $\epsilon$ plot
 for $k=6$, and $a=1.9$.
In Fig.~\ref{fig10}, $\sigma(n)$ is plotted for different values of $k$, for the case  $a=1.9$.
The synchronization behavior is comparable to that of a random network.
When $k > 8$ there is synchronization for $a=2.0$.  We checked our results with
higher values of $m$ also.  The results seem to converge for large system
sizes and to be independent of the time of evolution (size) of the
network.

\section{Conclusion}
We studied the spectrum of coupled map lattices and its relation to the 
stability properties of the spatially homogeneous solutions. We derived
conditions for the existence of such solutions using linear stability analysis. 
Conditions obtained from a global stability analysis are also provided. 
Our results are supplemented with numerical examples. For the numerical study
the quadratic (logistic) map is used for the site dynamics. We studied the
synchronization properties of coupled map lattices with different coupling 
topologies such as global coupling, nearest neighbor coupling, intermediate 
range coupling, random coupling, small-world coupling and real-world 
coupling. The coupling topology can crucially influence the synchronizability of
the CML. Our study can be generalized almost directly to other spatially 
extended systems.

\acknowledgments{
We thank Thomas Wennekers for the critical reading of the manusript.
}

\end{document}